\documentclass[11pt]{article}

\usepackage{graphics}

\begin{document}

\voffset 0.0cm
\setlength{\parindent}{5ex}
\newcommand{\be}{\begin{equation}}
\newcommand{\ee}{\end{equation}}
\newcommand{\ba}{\begin{eqnarray}}
\newcommand{\ea}{\end{eqnarray}}
\newcommand{\lb}{\label}
\newcommand{\bb}{\bibitem}
\newcommand{\dd}{\partial}
\newcommand{\half}{\frac{1}{2}}
\newcommand{\nn}{\nonumber}

 
\begin{center}

{ \huge 
Exact   General Relativistic  Disks\\

 with\\

\vspace{0.3cm}
 Magnetic Fields
 }

\vspace{1cm}
{\em  Patricio S. Letelier\footnote{e-mail: 
letelier@ime.unicamp.br} }

\vspace{0.3cm}
Departamento de Matem\'atica Aplicada-IMECC\\
 
Universidade Estadual de Campinas, 13083-970 Campinas. S.P., Brazil

\vspace{1cm}
 
{\bf \small Abstract}
\end{center}
The well-known  ``displace, cut, and reflect''  method used   to generate
cold disks from  given solutions of  Einstein equations is  extended to 
solutions of  Einstein-Maxwell equations.  Four exact solutions of the
these last  equations are used to construct models of hot disks with
surface density, azimuthal pressure, and azimuthal current. The solutions
are closely related to Kerr, Taub-NUT, Lynden-Bell-Pinault and to a
one-soliton solution.  We find that the presence of the magnetic field can 
change in a nontrivial way the different  properties of the  disks. In 
particular, the pure general relativistic instability  studied by
Bi\u{c}\'{a}k, Lynden-Bell and Katz [Phys. Rev. D47, 4334, 1993]  can be 
enhanced or cured by different distributions of currents inside the disk.
These  currents,  outside  the disk, generate  a variety of  axial symmetric 
magnetic fields. As  far as we know these are the first models  of hot disks
studied in the context of general relativity.

\vspace{0.4cm}


\newpage
\section{Introduction}

 The existent magnetic fields in the primeval cloud of plasma that
 originates   stellar objects like galaxies and 
 stars remain after the formation of  the objects. The magnetic 
lines of force
are said to be ``frozen in'' the fluid. 
This field  can play many different roles in astrophysics. Two
 examples
are:  The
 radius of a white dwarf for extreme relativistic degeneracy can change
 significantly  even when a  small quantity of magnetic energy is
 present \cite{shateu}. Binary and active-galactic-nuclei disks are hot
enough to be significantly or fully ionized, and their high conductivity
 allows  the current to  flow freely. The presence of magnetic fields
 strongly  influence the structure and evolution of these disks \cite{balhaw}.

Axially symmetric solutions of Einstein field equations 
corresponding to
disk like configurations of matter are of great astrophysical interest
and have been extensively studied. These solutions can be static or
stationary and with or without radial pressure.  Solutions for static
disks without radial pressure were first studied by Bonnor and Sackfield
\cite{BOSA}, and Morgan and Morgan \cite{MM1}, and with radial pressure
by Morgan and Morgan \cite{MM2}, and, in connection with gravitational
collapse, by Chamorro, Gregory and Stewart \cite{CHGS}. Disks with radial
 tension has been recently considered in \cite{gole}.

 The
   stability of  models  with no radial pressure can be
 explained by either assuming the
existence of hoop stresses or that the particles on the disk plane
move under the action of their own gravitational field in such a
way that as many particles move clockwise as counterclockwise. This
last interpretation is frequently made since it can be invoked to
mimic true rotational effects. A large class of thin disks
solutions were obtained by Letelier and Oliveira \cite{LO}.
 Solutions for self-similar static disks were
analyzed by Lynden-Bell and Pineault \cite{LP}, and Lemos \cite{LEM}. The
superposition of static disks with black holes were considered by
Lemos and Letelier \cite{LL1,LL2,LL3}, and Klein \cite{KLE}. Recently
Bi\u{c}\'{a}k, Lynden-Bell and Katz \cite{BLK} studied static disks as
sources of known vacuum spacetimes and Bi\u{c}\'{a}k, Lynden-Bell and
Pichon \cite{BLP} found an infinity number of new static solutions.

The above   mentioned solutions of the Einstein equations
 have no sources other than the disk itself, i.e., outside the disk 
their metrics are solutions of the vacuum Einstein equations
for static spacetimes with axial symmetry. The search  of new   solutions is
 facilitated in this case  by the fact that one of the Einstein  equations
 is equivalent to the  usual (linear)  Laplace equation.

In the present paper  we study models of disks that are characterized by a
 surface density, an azimuthal pressure and  a {\em current density}. Out 
side the disk  the metric satisfy the Einstein-Maxwell equations for a
 static axially symmetric spacetime filled with  a magnetic field. As
 far as we know these are the first models  of hot disks studied
 in the context of
 general relativity.

We find that the presence of the magnetic field can change in a nontrivial
 way the different  properties of the  disks. In particular, the pure general
 relativistic instability  studied by Bi\u{c}\'{a}k, Lynden-Bell and
 Katz \cite{BLK} can be enhanced of cured by different distributions
 of currents that  outside the disk generate magnetic fields of different
 forms.

In Sec. II we study the Einstein-Maxwell equations for a static axially
 symmetric spacetime. We introduce the well known 
``displace, cut, and reflect''  method
 to generate a disk from a given solution of these equations. We also
present the  expressions  for the main physical variables of the disk.
In the next section, Sec. III, four models of disks are studied.
 They are based  in simple solutions to the Einstein-Maxwell
 equations. The first two  are similar
 to Kerr and  Taub-NUT solutions. The third is a generalization of a
 Lynden-Bell and Pinault solution \cite{LP} and the last is one obtained using
solitonic techniques \cite{BZ}. In the last section , Sec. IV, we discuss
 some of the found  results and make some comments.

\section{Einstein-Maxwell Equations and Disks}

 The simplest  metric to describe  static
 and axially symmetric systems  is the Weyl metric,
\be
ds^2=fdt^2-f^{-1}[e^w(dz^2+dr^2)+r^2d\varphi^2] ,\lb{m1}
\ee
where the functions  $f$ and $w$ depend only on the cylindrical 
coordinates $r$ and $z$.

The Einstein-Maxwell system of equations is equivalent with,

\ba && G^{\mu\nu}=-T^{\mu\nu}, \lb{e1}\\
    && T^{\mu\nu}=F^{\mu}_{\;\;\alpha} F^{\alpha\nu}
 -\frac{1}{ 4 }g^{\mu \nu} 
F^{\alpha\beta}F_{\alpha\beta} , \lb{tel}\\
&&\nabla_\alpha F^{\alpha\mu}=0, \lb{max1}\\
&&F_{\mu\nu}=\dd_\mu A_\nu - \dd_\nu A_\mu, \lb{max2}
\ea
where all the symbols have their usual meaning.  Now we 
 choose the magnetic potential as
\be
A_\mu = A(r,z) \delta^{\varphi}_\mu,  \lb{A}
\ee 
in other words we take the  electromagnetic potential for a pure
 axially symmetric  magnetic field,
\be
 F_{\varphi z} = -\dd_z A , \;\;\; F_{r \varphi } = \dd_r A . \lb{fmna}
\ee
 Therefore  the physical components of the magnetic field are,
\ba
B_{\hat{r}}& =&-(f/r) e^{-w/2} \dd_z A \lb{Br},\\
B_{\hat{z}}& =&(f/r) e^{-w/2} \dd_r A \lb{Bz}.
\ea
From (\ref{m1})--(\ref{max1}) we find
the well known system of Einstein-Maxwell  equations for axially
 symmetric spacetimes with  pure magnetic fields \cite{KSM},
\ba
&&\dd_r w= \frac{r[(\dd_r f)^2-(\dd_z f)^2]}{2f^2}+
\frac{f[(\dd_r A)^2  -(\dd_z A)^2]}{r}, \lb{w1}\\
&& \dd_z w= \frac{r\dd_r f\dd_z f}{f^2}+\frac{2f\dd_r A \dd_z A}{r},
 \lb{w2}\\
 && \dd_r(r\dd_r f/f-fA\dd_r A/r)+\dd_z(r\dd_zf/f-fA\dd_z A/r)=0,
\lb{s1}\\
 &&\dd_r(f\dd_r A/r)+\dd_z(f\dd_z A/r)=0. \lb{s2}
\ea
Once the solution of the system of equations (\ref{s1})--(\ref{s2}) 
is known,  Eqs. (\ref{w1})--(\ref{w2}) give us the function $w$ as 
 a quadrature whose existence is guaranteed by the previous equations. 

The method that we shall use to generate the metric of the 
 disk,  as well as,  its 
material and electromagnetic content will be the well
 known ``displace, cut and
 reflect'' method that was first used in Newtonian gravity
 by Kusmin \cite{kus} and Toomre \cite{toom} and later
 in general relativity by most of  the authors that write about
 disks and also fine shells, defects, etc.

The  material content of the disk will be described  by functions that
 are distributions with
 support on the disk. The  method  that we shall use 
can be divided in the following steps:
 First, choose a 
surface that divide the usual space in two parts;  one with no 
singularities 
or  sources and the other with the sources. Second, disregard the
 part of
 the space with  singularities. Third, use the surface to make
 and inversion of  the nonsingular part of the space. The result
 will be
 a  space with a singularity that is a delta function with support
 on the surface.  This procedure is depicted in Fig. 1 for the
 gravitational  field of a punctual mass.

 In the present case the procedure is 
equivalent to make the transformation $z \rightarrow z_0+|z|$, with $z_0$
 constant. In
 the Einstein equation 
 for axial symmetry  we  have first  and second derivatives
 of $z$. Note that
$\dd_z|z|=2\theta(z) -1$ and $\dd_z \theta(z)=\delta(z)$ where
 $\theta(z)$ is
 the Heaveside function and $\delta(z)$ the usual Dirac 
distribution.
 
The procedure above mentioned applied to the metric (\ref{m1}) and 
then Einstein-Maxwell equations gives us 
\ba
&&G_{\mu\nu} =-(T^{elm}_{\mu\nu}+\hat{T}_{\mu\nu}),   \\
&&\nabla_{\mu} F^{\mu\nu}=\hat{J}^\nu,
\ea
with
\ba 
&&\hat{T}^{0}_0=\rho=e^{-w}(2\dd_z f -f\dd_z w)\delta(z) , \\
&&\hat{T}^{\varphi}_{\varphi}=P_\varphi=e^{-w}f\dd_z w \delta(z)  , \\
&&\hat{J}^\varphi=2f \dd_z A \delta(z)/r,
\ea
where  $\dd_z f$  should be understood as $\dd_z f|_{z=0+},$ etc.
 $T^{elm}_{\mu\nu}$ is the electromagnetic energy-momentum tensor defined 
in 
(\ref{tel}),  $\hat{T}_{\mu\nu}$ is the  matter  energy-momentum 
tensor on the plane $z=0$ and  $\hat{J}^\mu$ is the  current density
 on the plane $z=0$.  The ``physical meassure'' of length in the 
direction $\dd_z$ for the metric (\ref{m1}) is $(e^w/f)^{1/2} d z$,
then the invariant distribution is $\delta(z)/(e^w/f)^{1/2}$. Therefore
the surface density and the planar pressure or tension along 
the direction $\dd_\varphi$  are,
\ba 
&&\sigma= [(e^{-w}/f)^{1/2}(2\dd_z f -f\dd_z w)]_{z=0+}, \lb{sigma}\\
&& p_\varphi= [(fe^{-w})^{1/2} \dd_z w]_{z=0+}. \lb{pphi}
\ea
In the same form one can compute the tetradical component of
 the planar current density,
we find,
\be
j_\varphi =2 [ e^{w/2} \dd_z A]_{z=0+}.  \lb{j}
\ee
Another physically meaningful  quantity associated to non rotating disks
is the counter-rotation velocity,
\be
\left. V^2=p_\varphi/\sigma =\frac{f \dd_z w}{2\dd_z f -f \dd_zw}
 \right|_{z=0+} .
\ee
The disk is supposed to be formed by the same number of particles moving 
in circles  in one direction than in the opposite direction.
The specific angular-momentum of a particle with rest mass $\mu$ 
rotating at a radius r, defined as $h=\pi_\varphi/\mu =
g_{\varphi\varphi}d\varphi/ds$, is
\be
h=\frac{r\sqrt{p_{\varphi}\sigma}}{f\sqrt{\sigma^2-p_{\varphi}^2}}. \lb{h}
\ee
This last quantity can be used to detect instability of the disk against
 radial perturbations (ring formation). The condition of stability 
is usually
presented as 
\be
\frac{d ( h^2)}{dr}>0, \lb{dh}
\ee
i.e, for $ h>0$ we have stability when  the specific angular momentum  
is an increasing function of $r$ for $0<r<R$
 ($R$ the radius of the disk). This criteria is an extension of  Rayleigh
criteria of stability of a fluid at rest in a gravitational field, see
 for instance
\cite{LLfluids} . For other significant quantities 
for relativistic disks see \cite{BLK}. 
 
As we said before in the present case  the spacetime
 is static, hence in  the disk we
do not have a net movement of matter in one direction, but we have
 current along the $\dd_\varphi$ direction. This means that we
 have a kind of moving massless charge approximation. 

The field lines of the magnetic field are given by the ordinary differential
 equation,
\be
\frac{dz}{B_{\hat{z}}}=\frac{dr}{B_{\hat{r}}}.
\ee
From (\ref{Br}) and (\ref{Bz}) we find,
\be
\dd_r A dr+\dd_z A dz=0.
\ee
Thus the equation  $A(r,z)= C$ ($C$ constant) gives as the lines
of force  of the magnetic field.

\section{Exact Disks with Magnetic  Fields}

There are many techniques to find exact solutions of 
(\ref{s1})--(\ref{s2}), they are based in the fact that this system of
equations  has
extra symmetries (Geroch group \cite{ger}) and that is closely related 
to the principal sigma model \cite{BZ}.
In particular, we can write (\ref{s1})--(\ref{s2}) in the matrix form
\be 
\dd_r[r\dd_r(M)M^{-1})]+\dd_z[r\dd_z(M)M^{-1})]=0,\lb{Me} 
\ee
where
\be
M=\left( \begin{array}{cc}
 f^{1/2} & (f/2)^{1/2}A \\
(f/2)^{1/2}A & \frac{2r^2+fA^2}{2(f)^{1/2}}
\end{array} \right) ,
\lb{Md}
\ee
as well as (\ref{w1})--(\ref{w2}),
\ba
 \dd_r[w+\ln(r^4/f^2)]& =&\frac{1}{r}Tr[((\dd_r M)M^{-1})^2- 
 ((\dd_z M)M^{-1})^2], \lb{Mw1}\\
 \dd_z[w+\ln(r^4/f^2)]& =&\frac{2}{r}Tr[(\dd_r M)M^{-1}
)(\dd_z M)M^{-1})].   \lb{Mw2}
 \ea 
Note that the matrix $M$ is symmetric and $\det( M)=r^2$. Eq.
 (\ref{Me}) is  mathematical equivalent to the Ernst equation, 
although the physical and geometric content of these equations
 are  different \cite{KSM}.
Now we shall study models of disks based in
 simple solutions of (\ref{Me})\\

\vspace{0.3cm}
{\em A.  Disk Associated with a   Magnetic Dipole Solution }\\
\vspace{0.3cm}

A  Kerr like solution to (\ref{Me}) is
\ba  
f^{1/2}& =& \frac{p^2 x^2-q^2 y^2 -1}{(px+1)^2-q^2y^2} , \lb{fK}\\
A&=&-2\sqrt{2}pq\frac{p(x^2-y^2)+x(1-y^2)}{p^2x^2-q^2y^2-1}, \lb{AK}
\ea
where $x$ and $y$ are the spheroidal coordinates
\ba
2p x=\sqrt{r^2+(z+p)^2}+\sqrt{r^2+(z-p)^2}, \lb{x}\\
2p y=\sqrt{r^2+(z+p)^2}-\sqrt{ r^2+(z-p)^2}, \lb{y}
\ea 
 $p$ and $q$ are two constants related by
\be 
q^2=p^2-1. \lb{pq}
\ee
The function $w$ is given by
\be 
e^w=f^2 F^4 ,\lb{wF}
\ee 
with 
\be
F=\frac{p^2x^2-q^2y^2-1}{p^2((px+1)^2-q^2y^2) } .\lb{FK}
\ee
A similar solution, but in  a different system of coordinates,
 was first studied 
by Bonnor \cite{bondip}, for $p=1$ reduces
to the Weyl $\gamma$-metric  (with $\gamma=2$). This solution can be 
interpreted as a Weyl $\gamma$-metric  with a magnetic dipole located
in the origin of the coordinate system.

Now we perform the ``displace, cut and
 reflect'' method above mentioned, i.e., we do $z\rightarrow z_0+|z|$ in the
Kerr like  solution. This is equivalent to change only the
 definitions of the  coordinates $x$ and $y$ by the new definition 
\ba
2p x=\sqrt{r^2+(|z|+z_0+p)^2}+\sqrt{r^2+(|z|+z_0-p)^2}, \lb{xd}\\
2p y=\sqrt{r^2+(|z|+z_0+p)^2}-\sqrt{r^2+(|z|+z_0-p)^2}. \lb{yd}
\ea 

Now, the computation of the physical quantities like $\sigma, p_\varphi,$ etc. 
associated to this disk is straight forward, but the result is rather
cumbersome, 

\ba
\sigma&=&
8 [((2 \bar{y} z_{0} - 3) \bar{x}^2 - \bar{y}^2 - (\bar{x}^2 + \bar{y}^2)
 p \bar{x}) p^2 \bar{y} + (\bar{x}^2 + \bar{y}^2 - 2 q^
2 \bar{x}^2 \bar{y}^2) z_{0}  \nn \\
&& + ((3 \bar{y} z_{0} - 2) \bar{y} + \bar{x}^2 z_{0} + (\bar{x}^2 
+ \bar{y}^2) q^2 \bar{y}) p \bar{x}]  p^2/ 
 \nn\\
&&  
[ (q^2 \bar{y}^2 + 1 - p^2 \bar{x}^2) (q \bar{y} +
  1 + p \bar{x})^2 (q \bar{y} - 1 - p \bar{x})^2], \lb{sigkerr}\\
p_\varphi&=&
8(((\bar{x}^2 + \bar{y}^2)q^2 + 2)p\bar{x}\bar{y} - (\bar{x}^2 + \bar{y}^2
 + 2q^2\bar{x}^2\bar{y}^2)z_{0} - ((\bar{x}^
2 + \bar{y}^2)p  -\nn\\
&& 2\bar{x}\bar{y}z_{0})p^2\bar{x}\bar{y})p^2/[(q^2\bar{y}^2
 + 1 - p^2\bar{x}^2)^2(q\bar{y} + 1 + p\bar{x})(q\bar{y} 
- 1 - p\bar{x})],\nn\\
&&\lb{pkerr}\\
j_\varphi&=&
 - 4\sqrt{2}[(((3\bar{y}^2 - 1)\bar{x}^2z_{0} + 4\bar{y} +
 2(\bar{x}^2 + \bar{y}^2)q^2\bar{y})p\bar{x} -\nn\\
&& ((\bar{y}^2 + 2\bar{y}z_{0} - 1)\bar{y}   - 2(\bar{y} - z_{0})\bar{x}^2 +
 (4\bar{x}^2\bar{y}z_{0} - 2\bar{x}^2 +\nn\\
&& \bar{y}^4 - \bar{y}^2)q^2\bar{y}))p  - ((\bar{y}^2 + 1 - (\bar{y}^2
 - 3)q^2\bar{y}^2)z_{0} + 2(\bar{x}^2 + \bar{y}^2)p^4\bar{y}\nn\\ 
&& + (\bar{y}^2 - 4\bar{y}z_{0} - 1 
+ 2\bar{x}^2)p^3\bar{x}\bar{y})\bar{x}]q/(\bar{x}^2 - \bar{y}^2)^3,\lb{jker}
\ea
where
\ba
2p \bar{x}=\sqrt{r^2+(z_0+p)^2}+\sqrt{r^2+(z_0-p)^2}, \lb{x0d}\\
2p\bar{ y}=\sqrt{r^2+(z_0+p)^2}-\sqrt{r^2+(z_0-p)^2}. \lb{y0d}
\ea 
The magnetic potential $A$ is  given by the same function (\ref{AK}) but 
with $x$ and $y$ given by (\ref{xd}) and (\ref{yd}).

We shall  study the above functions   graphically.
In Fig. 2 we present the surface density $\sigma$ as a function of
 the radial coordinate $r$  for $z_0=2$ and different values of the
 parameter $q$ that controls the magnetic field.

The   density have a finite central maximum that decays fast. 
  The curve with the lowest maximum  has    $q=0$   and  
 the next  three curves have  $q= 0.5, 1,$ and $ 1.5$. The curve with $q=0$ 
represents the density of a disk with no
 magnetic field, i.e., the density of  a usual  cold disk.
We see that the inclusion of the magnetic field increases
 the  central density for the values of $q$ shown. Really, the central
 density  increases up to a value close to $2.5$ for $q=1.73$ and 
for values of $q\geq1.73$  decreases very fast and take
  negative values (not shown in the figure).

 The azimuthal 
pressure  is shown in Fig. 3 for $z_0=2$ and $q=0$  (bottom curve), $ 0.5,
 1, 1.5$ (top curve).  We have
that the pressure has a maximum that increases with $q$ and moves
 toward the
disk centre. For $q\geq 1.5$ we have similar behavior.

The azimuthal current is presented in Fig. 4 for the same values of
 the parameters $z_0$ and $q$ . The curve that
 corresponds to $q=0$ is the 
axis $r$. The  current has  similar  behavior than the pressure. 
We see that the behavior of the current and the pressure
is consistent with the presence of a ring-like  current of massless
 particles.

In Fig. 5 we  present the magnetic  potential $A$ and its  level curves that
represent the  the magnetic field for $z_0=2$ and $q=1$. We see that 
the form of the magnetic field also  supports the idea of a ring-like current
 distribution.

 Finally for this disk, we shall examine the stability against
 radial perturbations (ring formation). In Fig. 6 we present the radial
 dependence of the angular momentum $h$ for a cold disk $q=0$ for 
  $z_0=2$ (top curve), $z_0=2.5, 4, 3.4$ and $4$ (bottom curve). We see the
 the
disk with $z_0=2$ is unstable, h is not  an increasing
 function of $r$ for $0 \leq r \leq 14$. This
is a pure general relativistic instability. Now we shall study the
 stability of the disk with $z_0=2.5$ when we  switch on the magnetic 
field.

The radial 
 dependence of the angular momentum $h$ is presented in Fig. 7 for $q=0$
 (bottom curve), $0.5, 1, 1.5$ (top curve).
 Therefore  the disk with $q=1.5$ is unstable. Again the 
magnetic field introduces a pure gravitational instability. Note that
 the disk  with $q=0$ is stable and the inclusion of the magnetic field 
makes it gravitationally  unstable.  The density, pressure and current 
for this disk  look  similar than the ones for $z_0=2$, we shall
 comeback to this point later.\\

\vspace{0.3cm}
{\em B. A disk associated with  Taub-NUT like charge}\\
\vspace{0.3cm}

A Taub-NUT like solution is,
\ba
&& f^{1/2}=\frac{p^2(x^2-1)}{px^2+2x+p}, \lb{fTNUT} \\
 &&   A=2\sqrt{2}b y/p, \lb{ATNUT} \\
&&F=\frac{px^2+2x+p}{p(x^2-y^2)},
\ea
with 
\be
b^2=1-p^2. \lb{bp}
\ee
The function $w$ is given by (\ref{wF}), $b$ is a constant
 and $x$ and $y$
 are the same coordinates defined before.
This solution for $p=1$ also  reduces
to the Weyl $\gamma$-metric  (with $\gamma=2$).

As in the precedent case we find, 
\ba
\sigma&=&
 - 8[((\bar{y}^2 + 2 + \bar{x}^2)p^2\bar{y} - (\bar{x}^2 +
 3\bar{y}^2)z_{0})\bar{x} - ((2\bar{y}^2z_{0} - 3\bar{y} + \nn\\
&& z_{0})\bar{x}^2 - (\bar{y} - z_{0})\bar{y}^2)p]/[((\bar{x}^2 + 1)p +
 2\bar{x})^2(\bar{x^2} - 1)p], \lb{signut}\\
 p_\varphi&=&\frac
{ - 8[((2\bar{y}^2 - 1)\bar{x}^2 - \bar{y}^2)z_{0} - (\bar{y}^2 - 2
 + \bar{x}^2)p\bar{x}\bar{y}]}{((\bar{x}^2 + 1)p 
+ 2\bar{x})(\bar{x}^2 - 1)^2p},\lb{pnut}\\
j_\varphi&=&4\sqrt{2}(p\bar{x} - \bar{y}z_{0})(\bar{x}^2
 - 1)^2b/[(\bar{x}^2 - \bar{y}^2)^3p^2]. \lb{jnut}
\ea
The magnetic potential $A$ is the given by the same function (\ref{ATNUT}) but 
with $x$ and $y$ given by (\ref{xd}) and (\ref{yd}).

The surface density as a function of $r$ is presented in Fig. 8 for
  $b=0$ (bottom curve), $b=0.3, 0.6, 0.9$ (top curve) and $z_0=2$. We see
 that the 
  the density maximum increases when  the magnetic parameter $b$ increases.

We also see (Fig. 9),  for the same value of the parameters,
 a similar behavior for the current. The curve  $b=0$ is the  axis r. 
The pressure behaves
 in the opposite way  (Fig. 10),  it decreases when the magnetic parameter
 $b$ increases. For the  top and bottom curves we have $b=0$
 and $b=0.9$ respectively.

The field lines of magnetic field in this case are  quite simple, they
 are just
parts of hyperbola  branches  that cut the disk in an axially symmetric way, 
the axis of the disk being the symmetry axis.
Note that the distribution of current  is quite
 different in this case than in the precedent one. Now we have
 a more homogeneous distribution.

The angular momentum $h$ as a function
 of $r$ is shown in Fig. 11 for $z_0$=2 and
$b=0$ (top curve), $b=0.3, 0.6, 0.9$ (bottom curve). We see that the inclusion
of the magnetic parameter can stabilize the disk.  We shall
 comeback to this point later.\\

{\em C. A generalization of a  Lynden-Bell-Pinault  solution}\\
\vspace{0.3cm}

A magnetized Lynden Bell-Pinault  like solution is 
\ba
f^{1/2}&=&a^2/f_0 +c^2 r^2/f_0, \lb{fLBPm}\\
A&=&-\sqrt{2}\frac{abf_0+ dcr ^2}{a^2 _0+c^2r^2} \lb{ALBP}\\
w&=&w_0+2\ln(f/f_0) -8\ln a,   \lb{wLBPm}   
\ea where $f_0$ and $w_0$ are the respective functions associated to 
the Lynden-Bell-Pinault solution \cite{LP}:  
\ba 
f_0&=&[(z+\sqrt{r^2+z^2})/2]^{2n}, \lb{fLBP}\\
w_0&=&4n^2\ln \frac{z+\sqrt{r^2+z^2}}{2\sqrt{r^2+z^2}},\lb{wLBP}
\ea
  $a,b,c,d$ and $n$ are constants  restricted by
  the relations $ad-cb=\pm 1$, and $n\leq 1$. This solution was found by
 applying the transformation $M=B^{-1}M_0 B$ with 
\be
B= \left( \begin{array}{cc} a& b\\ c&d \end{array}\right). \lb{B}
\ee
 $M_0$ being the matrix (\ref{Md}) com $A=0$. Note  the
 right hand side of
Eqs. (\ref{Mw1}) and  (\ref{Mw2}) are invariant under the
 transformation 
 $M=B^{-1}M_0 B$.

Now doing $z\rightarrow |z|+z_0$ in the previous solution we find
after doing $z_0=0$,

\ba
&&\sigma=
 r^{3n}2^{2n^2-n+2}(1-n)a^{4}n/[(r^{2n}a^2 + 2^{2n}c^2
r^2)r] \lb{sigLBP}\\
&&p_\varphi=r^{3n-1}2^{2n^2-n+2}\frac{[ r^{2n}a^{2}n - 2^{2n}(2-n)c^2r^2  ]
 a^4 
n}{(r^{2n}a^2 + 2^{2n}c^2r^2)^2}  \lb{pLBP}\\
&& j_\varphi=\pm  2^{2n+5/2-2n^2}cnr^{1-2n}/a^3.\lb{jLBP}
 \ea
Note that this solution has well behave density that is singular
 in the center of the disk, like the original LBP solution. The pressure
is positive in the center,  but for $r>r_c$ it   becomes negative, 
\be
r_c=\left[\frac{n a^2}{2^{2n}(2-n) c^2}\right]^{1/(2-2n)}\lb{rc},
\ee
i.e., we have  tensions.
This is  a general feature of  solutions generated in the same form
 using a Weyl  solution as a seed.

Hence in this case we have that  this particular magnetization of
 the LBP disk introduces tensions. \\

\vspace{0.3cm}
{\em D. A particular disk associated to a   1-soliton like solution}
\vspace{0.3cm}

A particular 1-soliton  solution is
 
\ba
 && f^{1/2}=\frac{(c_{0}^2+c_{1}^2)\mu r }{c_{0}^2\mu^2
-c_{1}^2 r^2}, \lb{f1sol}\\
&& A=-\sqrt{2}\frac{ c_0 c_1(\mu^2+r^2)}{(c_{0}^2+c_{1}^2)\mu},
 \lb{A1sol}\\
&& F=\frac{c_{0}^2\mu^2 -c_{1}^2 r^2}{\sqrt{r(c_{0}^2+c_{1}^2)}(\mu^2+
r^2)},
 \lb{F1sol}
\ea
where
\be 
\mu=z-z_0 +\sqrt{r^2+(z-z_0)^2}  \lb{mu}
\ee
and $c_0, c_1$ and $z_0$ are constants. This solution was
 obtained by applying
 the Belinsky-Zhakarov \cite{BZ} method with a single pole scattering 
matrix. A
 similar solution was studied in \cite{sol}

As before by doing $z\rightarrow |z|$ we find

\ba
\sigma&=&
16(c_0^2 + c_1^2)\mu^2 r^2/[(c_0 \mu + c_1 r)^2(c_0 \mu -
 c_1 r)^2],\lb{sigsol}\\
p_\varphi&=&8(\mu^2 - r^2))/[(c_0 \mu + c_1 r)(c_0\mu - c_1 r)]\lb{psol}
\\
j_\varphi&=&
4\sqrt{2}r\mu^2(\mu^2 - r^2) c_0 c_1 /(\mu^2 + r^2)^3 \lb{jsol}
\ea
For this solution we have that the central density of the disk is zero
 but not the pressure.  

In the  four solutions presented we have chosen the integrations
 constants associated with the  quadrature of $w$ in such a way
 that $w \rightarrow 0$ on the symmetry axis. Thus we  have a regular 
 symmetry axis  out side the sources.
There are many  known exact solutions of the Einstein-Maxwell
 equations with axial symmetry, as well as, many methods to find them.
In particular, the first two solutions presented  are
 specializations of a
 Kerr-NUT like solution or two soliton solution. We have presented them
 separated because of algebraic simplicity.

\section{ Discussion}

For the disks studied in the present paper we made a rather 
 complete study of  parameters,  we   only mentioned 
 the more relevant aspects of each solution. In particular, for  
solutions generated from
 finite sources like the one in Fig. 1, we have that the density of the disk
is a decreasing function of the ``cut'' parameter  $z_0$. For a Newtonian disk
 generated from Newton's law $\Phi \sim M/\sqrt{r^2+z^2}$
 the ``displace, cut,
 and reflect method'' gives the surface density $\sigma_N \sim M z_0/(r^2+
z_0^2)^{3/2}$ (Kusmin-Toomre disk).  Thus the central density $\sigma_N(0)\sim
 M/z_0^2$. For the first two studied disks A. and B.  we put the 
`` mass''  parameter equal to one. This is equivalent to have coordinates $t,r
, z$  dimensionless, one can recuperate this parameter by multiplying
 the line element by  $m$.  The numerical study of the two first disks 
shows that the central density  also decays with $z_0$, but  with 
a different law than  the Kusmin-Toomre disk.

We find that the magnetic field stabilize   the disk against
 radial perturbation when the distribution of current
 that generate the field has  its origin in a NUT like parameter. And 
made the disk unstable when  the distribution of current is originated by
a Kerr like  parameter.
In a different, context we studied a similar
 phenomena, we found that the inclusion of magnetic mass in a
 Weyl solution with unstable orbits (Schwarzschild solution  plus a
 dipolar halo) makes the orbits less  chaotic \cite{letvienut}, i.e., more
 stable. Also the  opposite is true for
particles moving in counter rotation in a slow rotating Kerr
 solution  plus a dipolar halo. The rotation makes the orbits more
 chaotic  \cite{letviekerr}.  These are pure general relativistic effects.

In the present work we studied simple four solutions of the Einstein-Maxwell 
equation generated by well known new-solution-generating
 algorithms \cite{KSM}. There are quite a few solutions of the Einstein
 Maxwell equations with axial symmetry that can be used to generate disks.
For instance  in the present context members of a Tomimatsu-Sato like
 family of solutions are
 good candidates. But the algebraic complexity involved in the finding of the
physical variables of the disks is not trivial.

The hot disks studied in this paper have not radial pressure nor the charges
 have  mass. There are several methods to add radial pressure 
to disks, they are
 based in the fact that a coordinate transformation that keep  $det M$
as a harmonic function in the variables $r,z$ does not change the form 
of the metric (\ref{m1}).

A second  degree of complexity is to add rotation to the disks, also suitable
solutions of the Einstein-Maxwell equations for stationary spacetimes are
 well known. Again the algebraic complexity is not trivial in this case.
The inclusion of rotations, in principle, can
 be used to introduce mass in the charge carriers. A  particular case of
hot rotating disk is under study.

Finally, we want to mention that all the computation of this work was 
 performed
using the algebraic programming system Reduce \cite{reduce}

\vspace{0.3cm}
\noindent
{\bf Acknowledgments}

I want to thank  CNPq and FAPESP for financial support. Also
 S.R.  Oliveira  for discussions.

\newpage

\begin{center}
\rotatebox{-90}{
\resizebox{40mm}{120mm}{%
\includegraphics{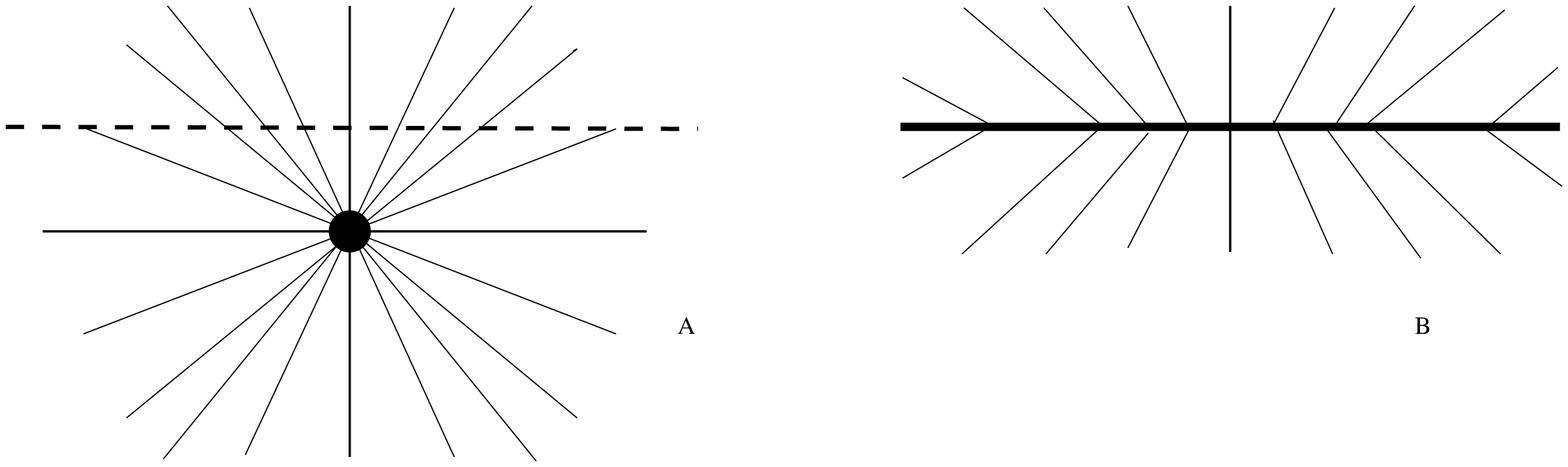}}}%
\end{center}
\vspace{0.3cm}
Fig. 1. The  illustration of  the ``displace, cut, and reflect''
 method for the generation of a disk from the gravitational field
 of a punctual mass. In  A we displace and cut  out the singularity
 with a plane, in  B we reflect  the field on the plan. We end up with
the field of a disk.
\vspace{1cm}

\begin{center}
\rotatebox{-90}{
\resizebox{70mm}{70mm}{%
\includegraphics{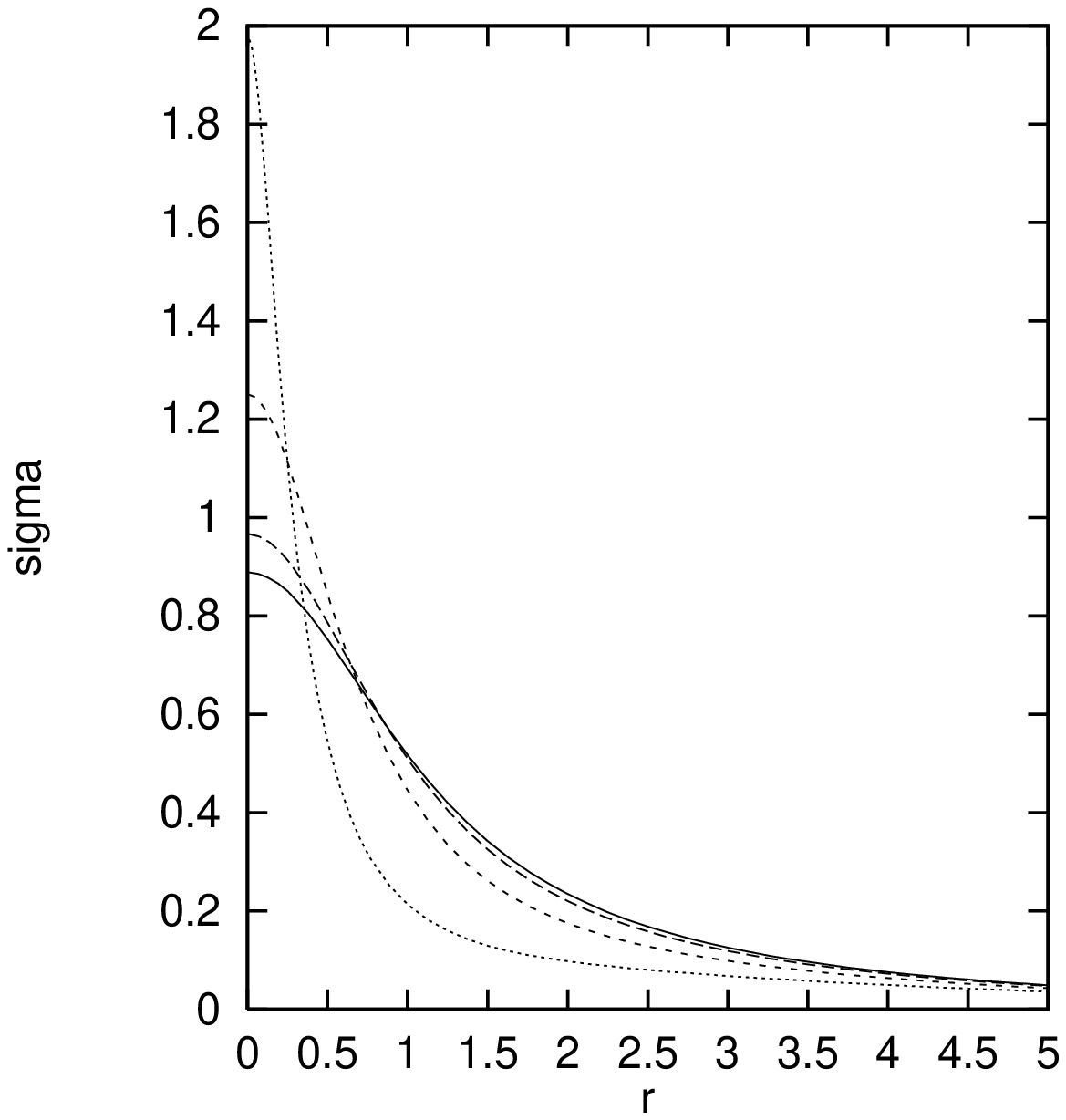}}}%
\end{center}
\vspace{0.3cm}
Fig. 2. The surface  density $\sigma$ as a function  of 
$r$  for $z_0=2$ for   $q=0 $ (lower maximum), $ 0.5,1,
 1.5$ (higher maximum). The parameter q controls the magnetic field. 
\

\begin{center}
\rotatebox{-90}{
\resizebox{70mm}{70mm}{%
\includegraphics{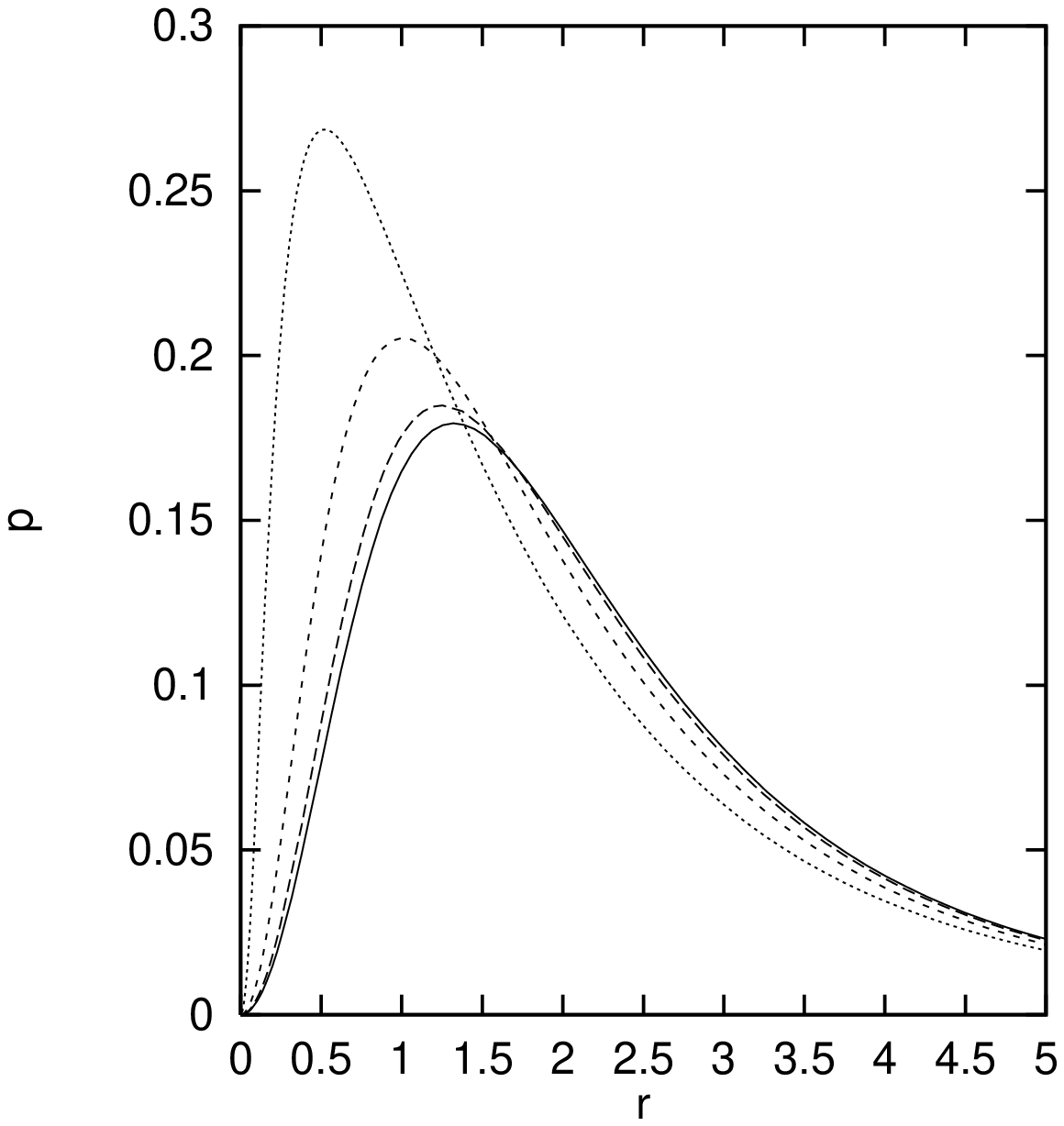}}}%
\end{center}
\vspace{0.3cm}
Fig. 3. The azimuthal 
pressure for $z_0=2$ and $q=0$  (bottom curve), $ 0.5,
 1, 1.5$ (top curve). We have
that the pressure has a maximum that increases with $q$ and moves
 toward the disk centre.

\begin{center}
\rotatebox{-90}{
\resizebox{70mm}{70mm}{%
\includegraphics{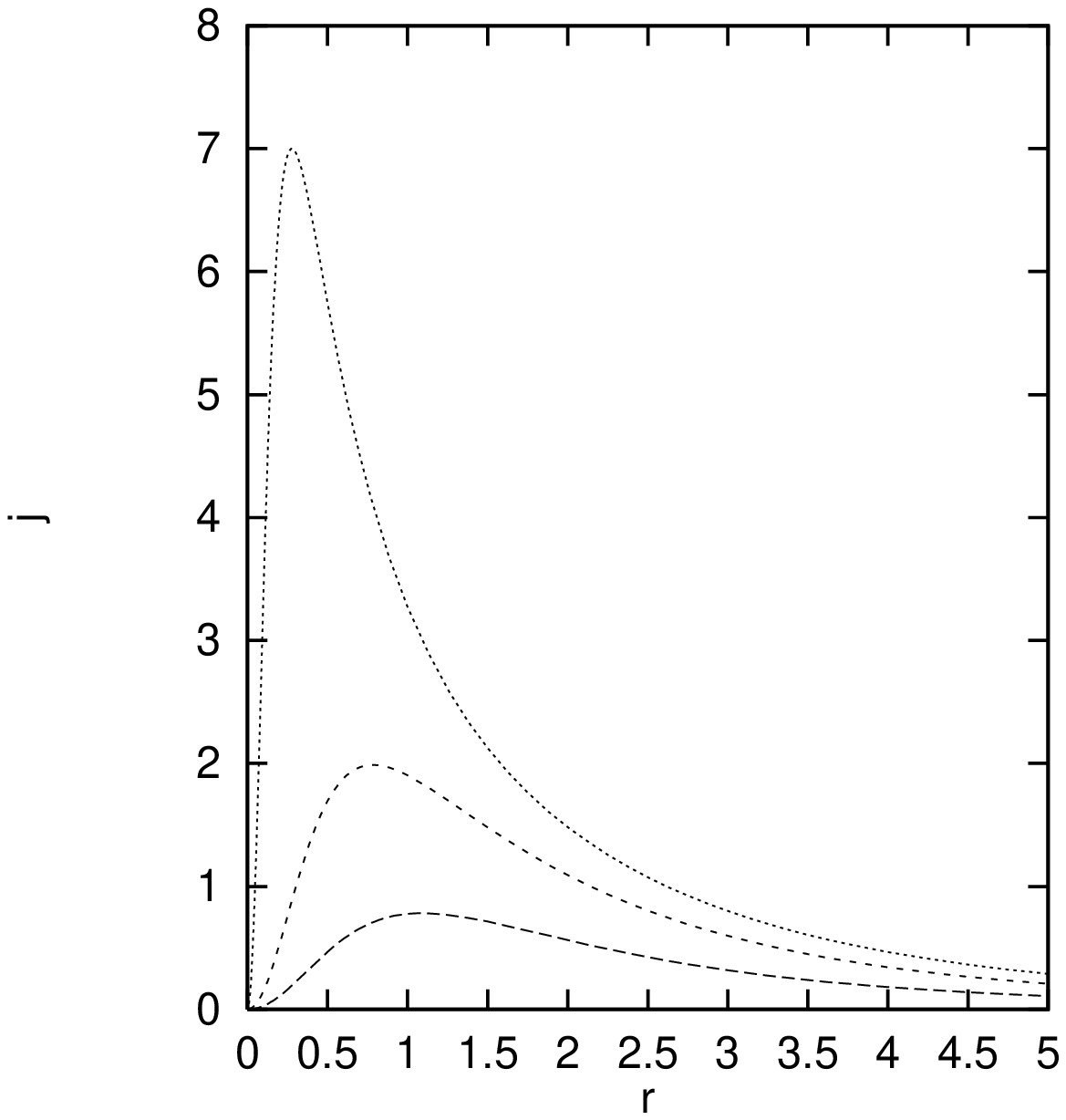}}}%
\end{center}
\vspace{0.3cm}
Fig. 4.  The azimuthal current  for the same values of
 the parameters $z_0$ and $q$. The curve that corresponds to $q=0$ is the 
axis $r$.
\vspace{0cm}

\begin{center}
\rotatebox{-90}{
\resizebox{100mm}{100mm}{%
\includegraphics{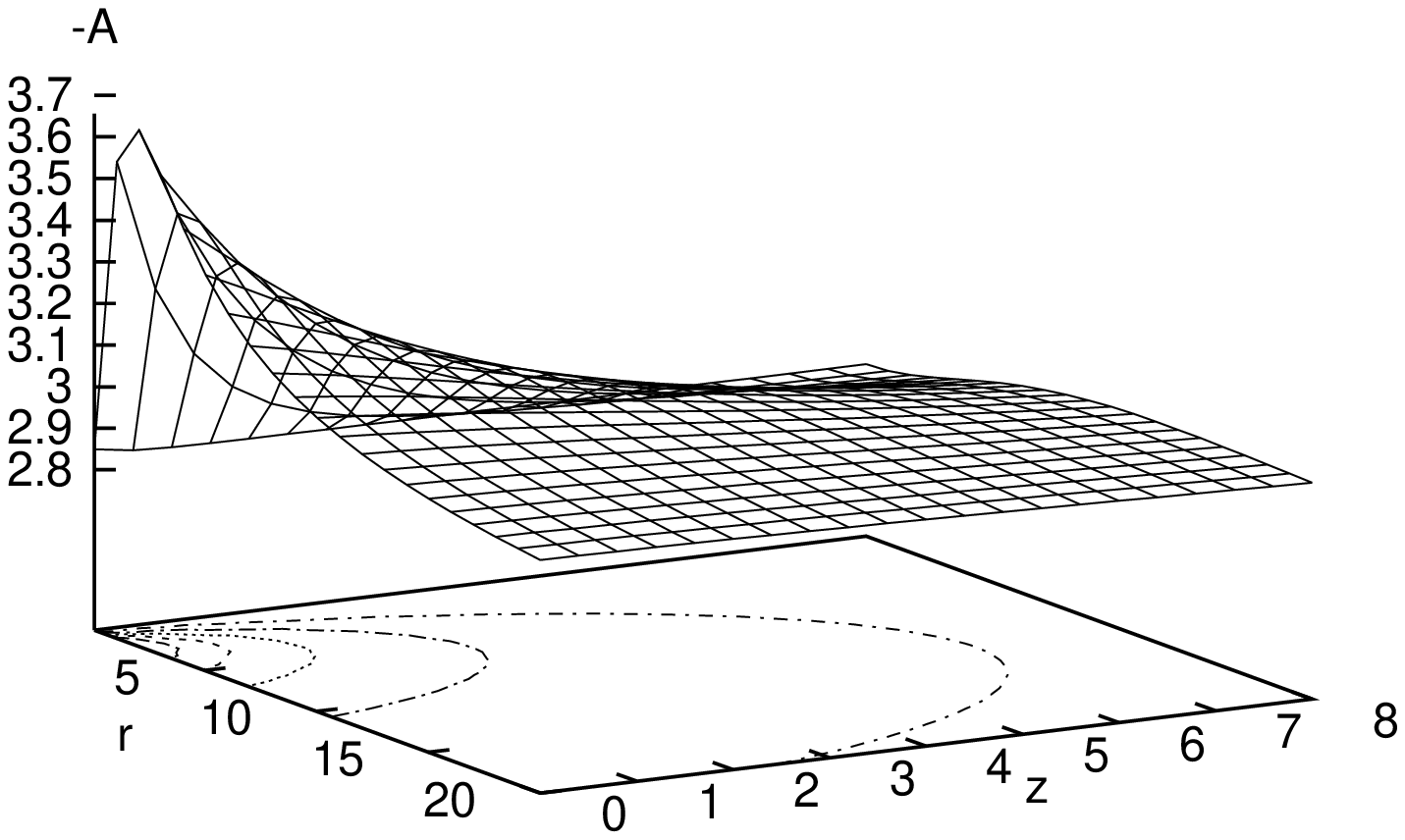}}}%
\end{center}
\vspace{0.3cm}
Fig. 5. The function  $A$ and its  level curves that
represent the  the magnetic field for $z_0=2$ and $q=1$. We see that 
the form of the magnetic field support the idea of a ring-like current
 distribution.
  \vspace{0cm}

\newpage
\begin{center}
\rotatebox{-90}{
\resizebox{70mm}{70mm}{%
\includegraphics{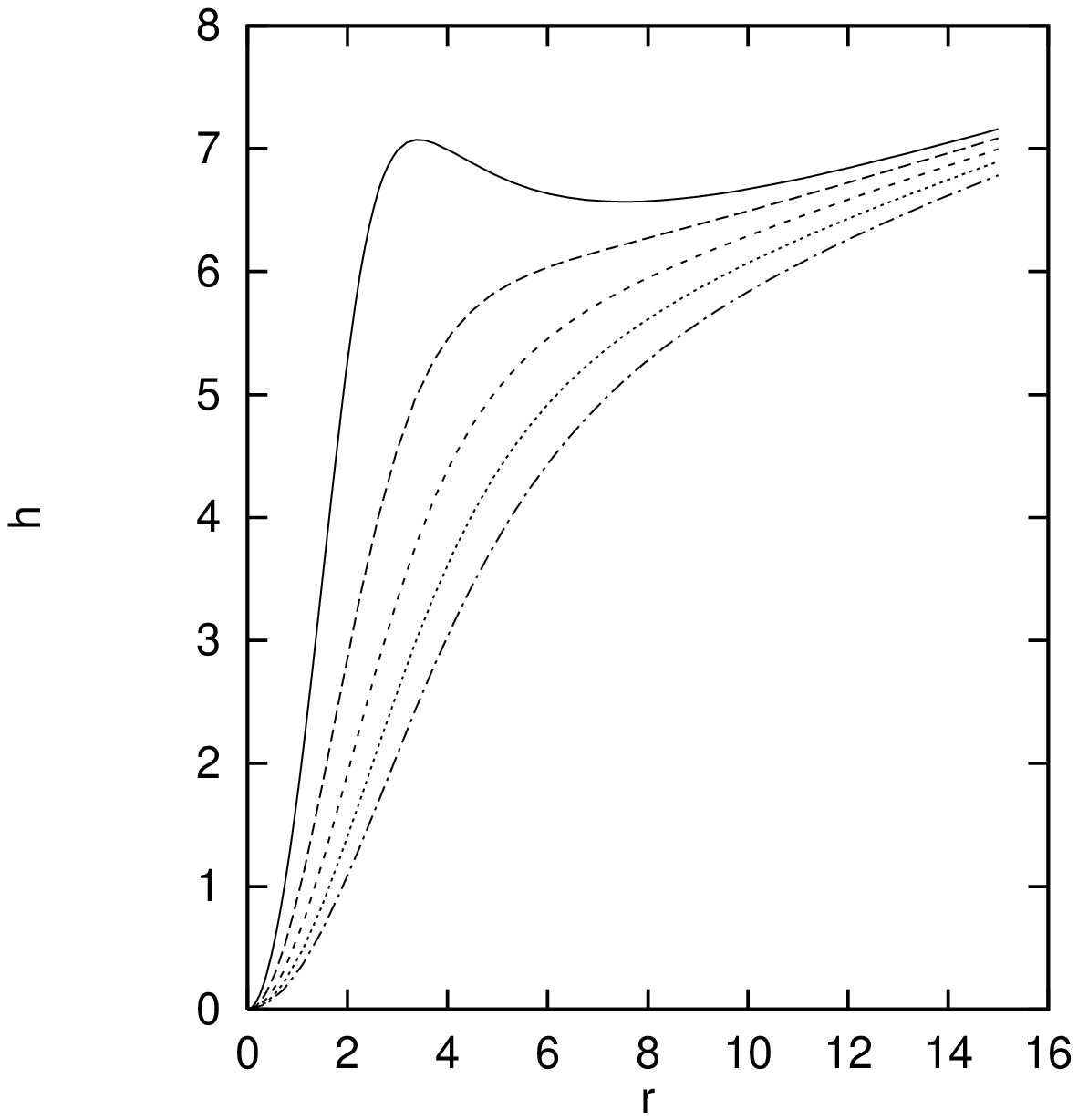}}}%
\end{center}
\vspace{0.3cm}
 Fig. 6.  The radial
 dependence of the angular momentum $h$ for a cold disk $q=0$ for 
  $z_0=2$ (top curve), $z_0=2.5, 4, 3.4$ and $4$ (bottom curve).
  \vspace{0cm}

\begin{center}
\rotatebox{-90}{
\resizebox{70mm}{70mm}{%
\includegraphics{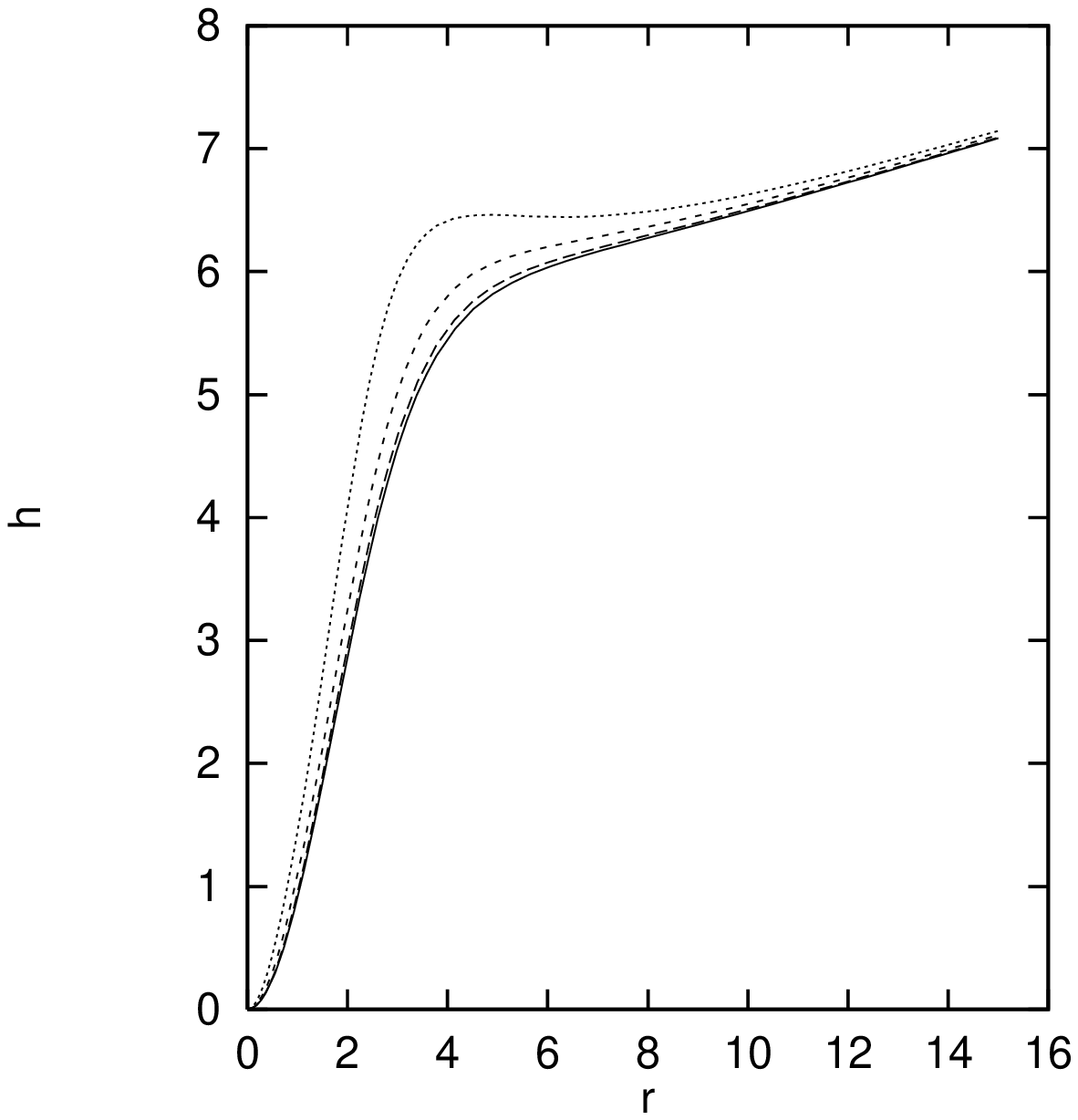}}}%
\end{center}
\vspace{0.3cm}
 Fig. 7. The radial
 dependence of the angular momentum $h$ for $q=0$ (bottom curve), $0.5, 1,
 1.5$ (top curve). In this case colder disks are more stable.
  \vspace{1cm}

\begin{center}
\rotatebox{-90}{
\resizebox{70mm}{70mm}{%
\includegraphics{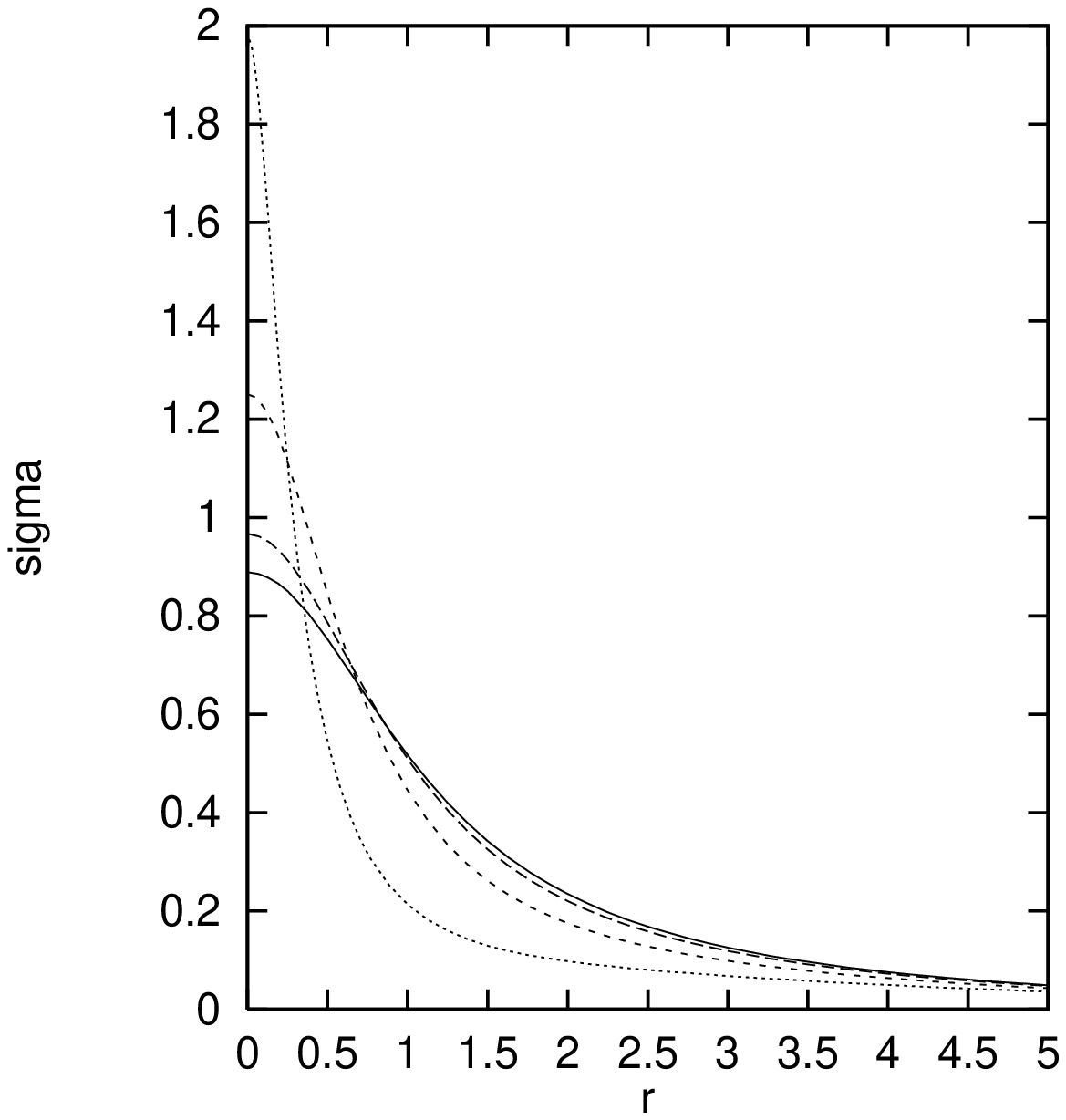}}}%
\end{center}
\vspace{0.3cm}
 Fig. 8. The surface density as a function of $r$ for
  $b=0$ (bottom curve), $b=0.3, 0.6, 0.9$ (top curve) and $z_0=2$.  The density
maximum  increases when  the magnetic parameter $b$ increases.
 \vspace{0.3cm}

\begin{center}
\rotatebox{-90}{
\resizebox{70mm}{70mm}{%
\includegraphics{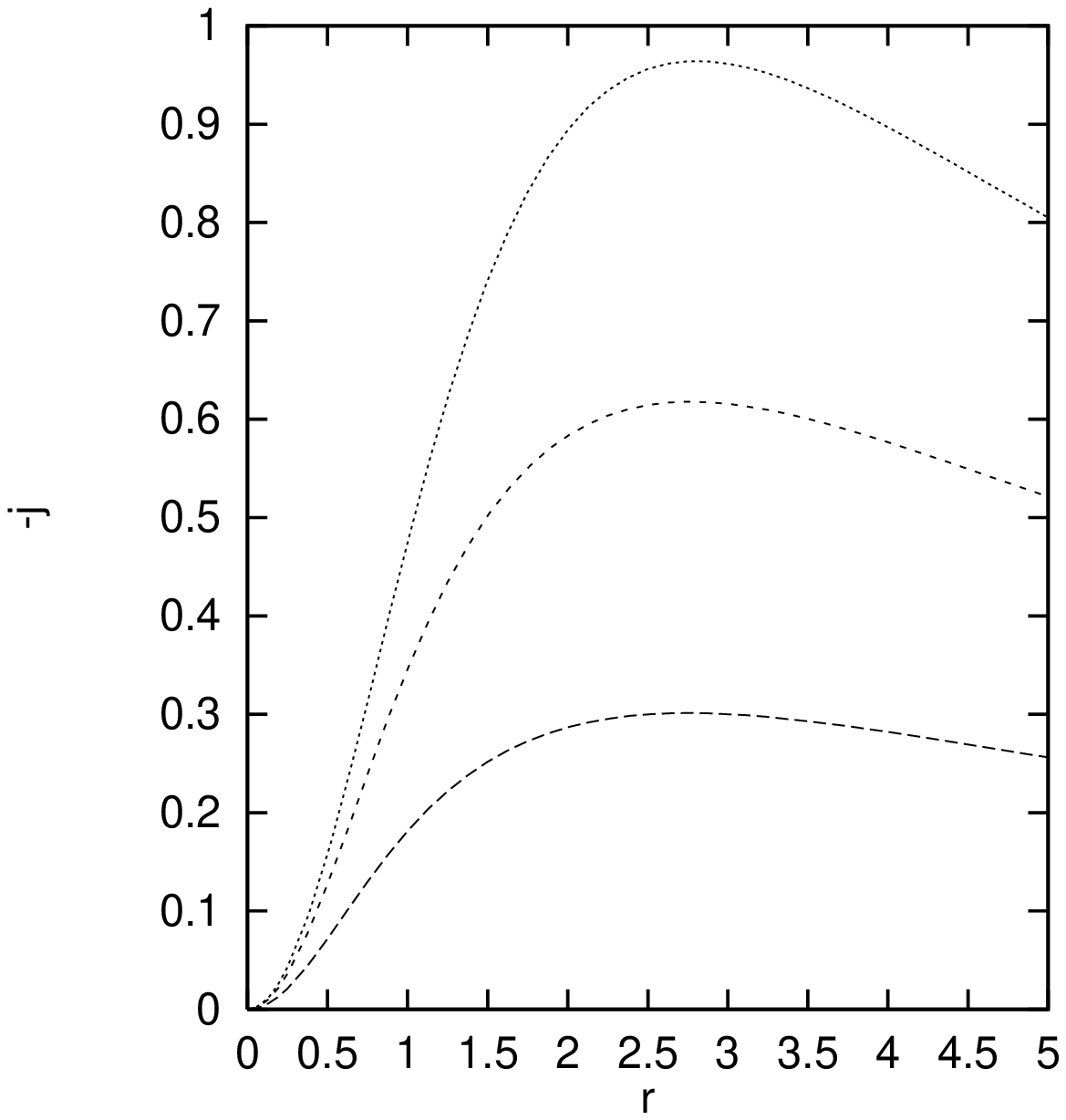}}}%
\end{center}
\vspace{0.3cm}
  Fig. 9. The current  for the same value of the parameters of Fig 8.
The curve  $b=0$ is the  axis r. The current
 and the density have   a similar behavior.
\vspace{0 cm}

\begin{center}
\rotatebox{-90}{
\resizebox{70mm}{70mm}{%
\includegraphics{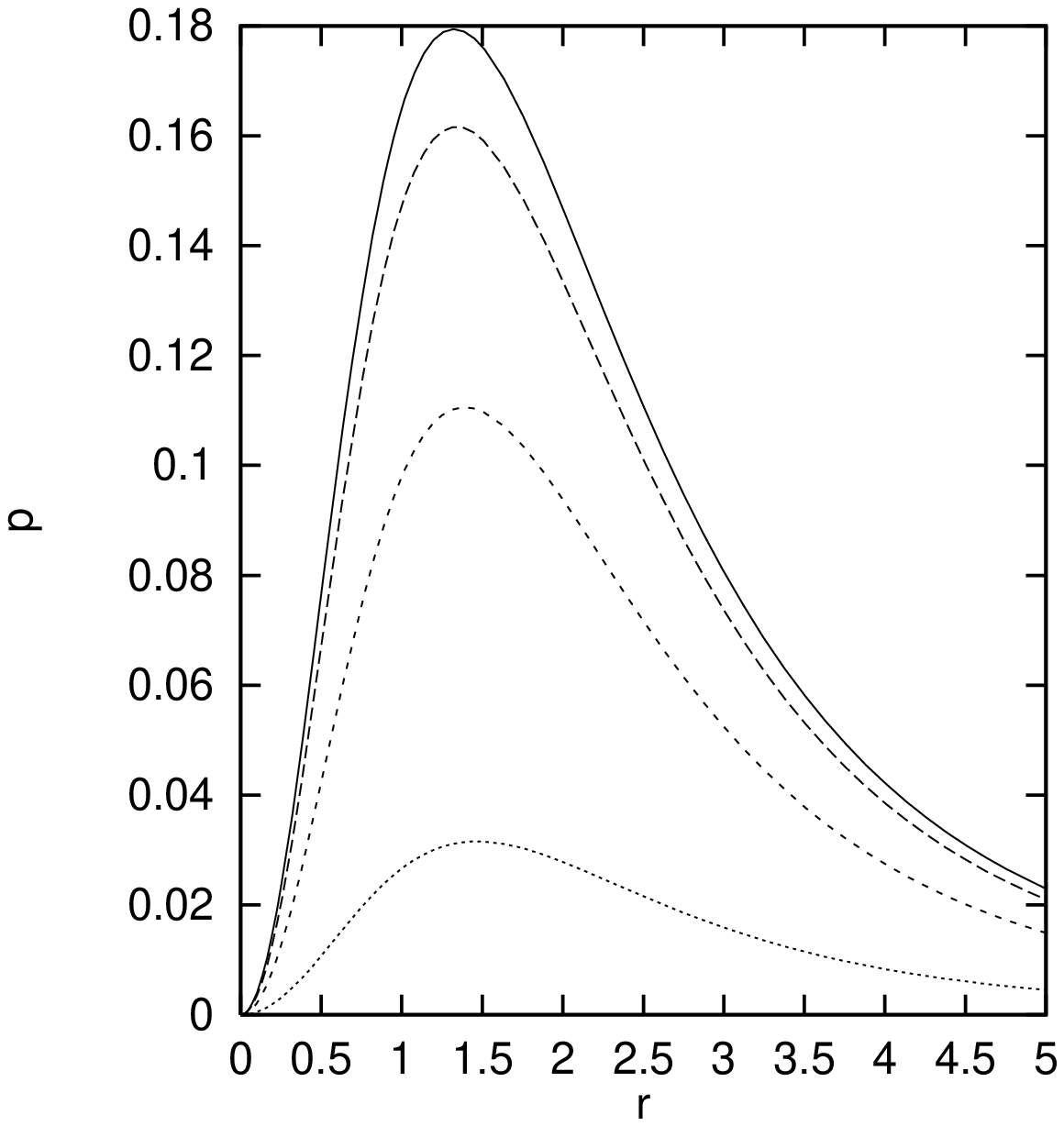}}}%
 \end{center}
\vspace{0.3cm}
Fig 10. The pressure  decreases
 when the magnetic parameter
 $b$ increases. For the  top and bottom curves we have $b=0$
 and $b=0.9$, respectively.
\vspace{1cm}

\begin{center}
\rotatebox{-90}{
\resizebox{70mm}{70mm}{%
\includegraphics{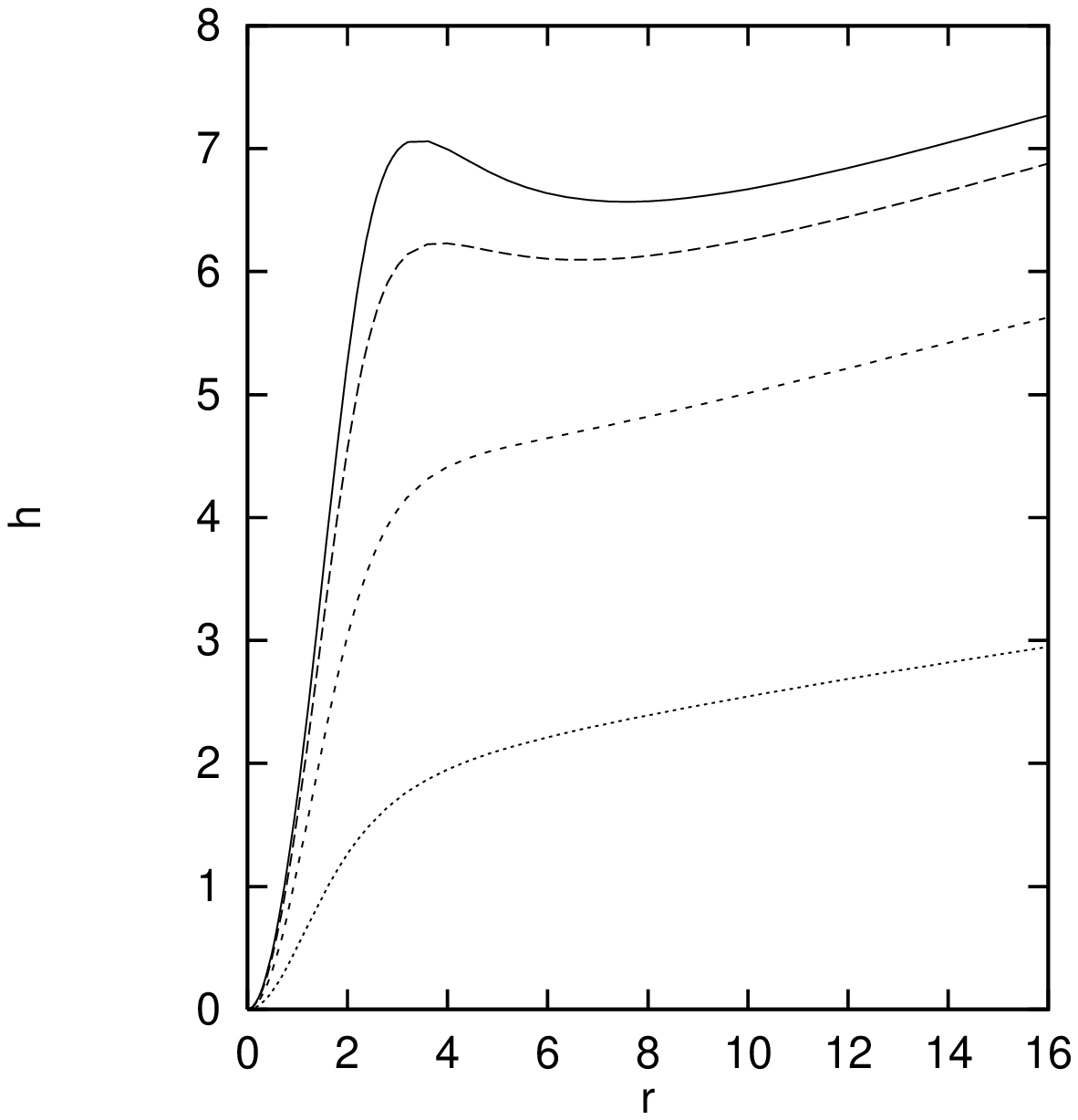}}}%
\end{center}
\vspace{0.3cm}
Fig. 11.   $h$ as a function
 of $r$ for $z_0$=2 and
$b=0$(top curve), $b=0.3, 0.6, 0.9$(bottom curve). In this case hoter disks
 are more stable.
\vspace{0.3cm}

\end{document}